\documentclass[onecolumn,11pt,msmath,amsmath,amssymb,aps]{revtex4-1}
\usepackage[colorlinks=true,urlcolor=blue,linkcolor=blue,citecolor=blue]{hyperref}
\usepackage{subfigure}
\usepackage{tablefootnote}
\usepackage{wrapfig}
\usepackage{graphicx}
\usepackage{textcomp}
\usepackage{bm}
\usepackage{xfrac}
\usepackage{color}
\usepackage{xcolor}
\usepackage{setspace}
\usepackage{silence}
\usepackage{epstopdf}
\newcommand{\BiTe} {{Bi}$_2${Te}$_3$}
\RequirePackage[normalem]{ulem} 
\RequirePackage{color}
\definecolor{RED}{rgb}{1,0,0}
\definecolor{BLUE}{rgb}{0,0,1} 

\newcommand{\mr}[1]{\mathrm{#1}}

\makeatletter
    \renewcommand\@make@capt@title[2]{%
     \@ifx@empty\float@link{\@firstofone}{\expandafter\href\expandafter{\float@link}}%
      {\textbf{#1}}\@caption@fignum@sep#2\quad}%
    \makeatother

\begin{document}
\title{Josephson Effect and Charge Distribution in Thin \BiTe\ Topological Insulators}

\author{M.P. Stehno}
\altaffiliation{Both authors contributed equally to this work}
\affiliation{Faculty of Science and Technology and MESA$^+$ Institute for Nanotechnology, University of Twente, 7500 AE Enschede, The Netherlands}
\affiliation{Physikalisches Institut EP3, University of W\"urzburg, Am Hubland, D-97070 W\"urzburg}

\author{P. Ngabonziza}
\altaffiliation{Both authors contributed equally to this work}
\affiliation{Faculty of Science and Technology and MESA$^+$ Institute for Nanotechnology, University of Twente, 7500 AE Enschede, The Netherlands}
\affiliation{Max Planck Institute for Solid State Research, D-70569 Stuttgart, Germany}
\affiliation{Department of Physics, University of Johannesburg, P.O. Box 524 Auckland Park 2006, Johannesburg, South Africa}

\author{H. Myoren}
\affiliation{Graduate School of Science and Engineering,
	Saitama University,
	255 Shimo-Okubo, Sakura-ku, Saitama 338-8570, Japan}

\author{A. Brinkman}
\affiliation{Faculty of Science and Technology and MESA$^+$ Institute for Nanotechnology, University of Twente, 7500 AE Enschede, The Netherlands}

\date{\today}
\begin{abstract}
Thin layers of topological insulator materials are quasi-two-dimensional systems featuring a complex interplay between quantum confinement and topological band structure.  {To understand the role of the spatial distribution of carriers in electrical transport,} we study the Josephson effect, magnetotransport, and weak anti-localization in bottom-gated thin \BiTe\  topological insulator films.
  We compare the experimental carrier densities to a model based on the solutions of the self-consistent Schr\"odinger-Poisson equations and find excellent agreement.  The modeling allows for a quantitative interpretation of the weak antilocalization correction to the conduction and of the critical current of Josephson junctions with weak links made from such films without any \textit{ad hoc} assumptions.   
\end{abstract}

\newpage
\maketitle
\section*{Introduction}
\paragraph*{}Three dimensional topological insulators (3D TIs) are a relatively new class of semiconductor materials with a band inversion in the bulk band structure.  The ordering of valence and conduction bands is reversed and surface states emerge that are protected by the topology of the band structure.\textsuperscript{\cite{LFu_CLKane_2007}}  These surface states feature a Dirac-like energy dispersion with a spin structure that is linked to the crystal direction ("spin-momentum-locking"). Topological quantum states are predicted to generate new low-energy effective modes of the electronic system, Majorana bound states, when topological surface states (TSS) are coupled to conventional \textit{s}-wave superconductors.\textsuperscript{\cite{fu_superconducting_2008,YTanaka_2009,APotter_2011}} The Majorana quasi-particle bound state in condensed matter systems could potentially be used as topological qubit to perform fault-tolerant computation.\textsuperscript{\cite{CNayak_2008,AStern_2010}} Recently, signatures of Majorana fermions have been found in quantum structures based on 2D and 3D TI Josephson devices.\textsuperscript{\cite{wiedenmann_4-periodic_2016, bocquillon_gapless_2017, deacon_josephson_2017}}

\paragraph*{}While \BiTe\  and other  Bi-based TIs have the advantage of significantly larger band gaps (several 100\,meV), and although proximity-induced superconductivity in Josephson devices fabricated on thin films of these materials have been reported;\textsuperscript{\cite{schuffelgen_boosting_2017,stehno_signature_2016,LGalletti_2017,PSchuffelgen_2019,PSchuffelgen_2019_01,SCharpentier_2017}}  little is known about the spatial distribution of carriers in normal state transport and induced supercurrent. A nonuniform charge distribution of intrinsic dopants\textsuperscript{\cite{MatthewBrahlek2015}} and extrinsic impurity contaminations\textsuperscript{\cite{PNgabonziza_2018,hoefer_intrinsic_2014}} are known to be present in Bi-based TI films, causing band bending close to interfaces.\textsuperscript{\cite{MSBahramy_2012,CChen_2012}} These effects also need to be taken into consideration in the discussions of proximity effects in these materials. Using thinner samples would minimize the contribution of bulk modes in electronic transport by decreasing the total number of dopant charges. However, when the sample thickness approaches the length scale of electrostatic screening or becomes comparable to the typical spreading of the TSS wave function into the bulk, the coupling between the superconductor and the TI material could change substantially.
\paragraph*{}To address above effects, we study electrical transport in thin \BiTe\ films with small, but finite, residual doping.  Unlike the compound Bi$_2$Se$_3$, the \BiTe\  material is not prone to having a large number of surface vacancies that lead to the formation of deep quantum wells at the surfaces.\textsuperscript{~\cite{PDCKing2011, MatthewBrahlek2015}}
This allows us to  observe the nontrivial interplay between band bending in the bulk and the surface states, resulting in a partial decoupling of the latter. We compare devices prepared from thin films of average thicknesses 6 nm and 15 nm. The choice of 6 nm for the thinner film is motivated by the objective of eliminating most of the TI bulk 
without opening a hybridization gap.\textsuperscript{\cite{liu_oscillatory_2010}} As we use a bottom gate to modulate electrical transport properties, the value for the thicker film is chosen to be comparable to the electrostatic screening length of the films (bulk screening length estimated to be in the range of 10 to 30 nm).\textsuperscript{\cite{jenkins_dirac_2013-1,MatthewBrahlek2015}}

\paragraph*{}First, we observe an unusual gate-voltage dependence of the critcal current in Josephson devices and find that it does not scale with the carrier densities obtained from Hall effect measurements.  Second, to understand the unusual gate-voltage dependence of transport properties, we calculate self-consistently the gate-dependent carrier distributions for a tight-binding model 
of the film and vary the doping levels and surface charges.   A uniform dopant distribution yields the best fit with the Hall effect data.  Lastly, backed by the theoretical model, we are able to interpret the magnitude of the observed weak antilocalization correction to the conductivity quantitatively and conclude that the critical current of the Josephson devices maps the change in shape of the (sub)band structure of the thin film when a gate voltage is applied. 
\begin{figure*}[!t]
	\includegraphics[width=0.8\textwidth]{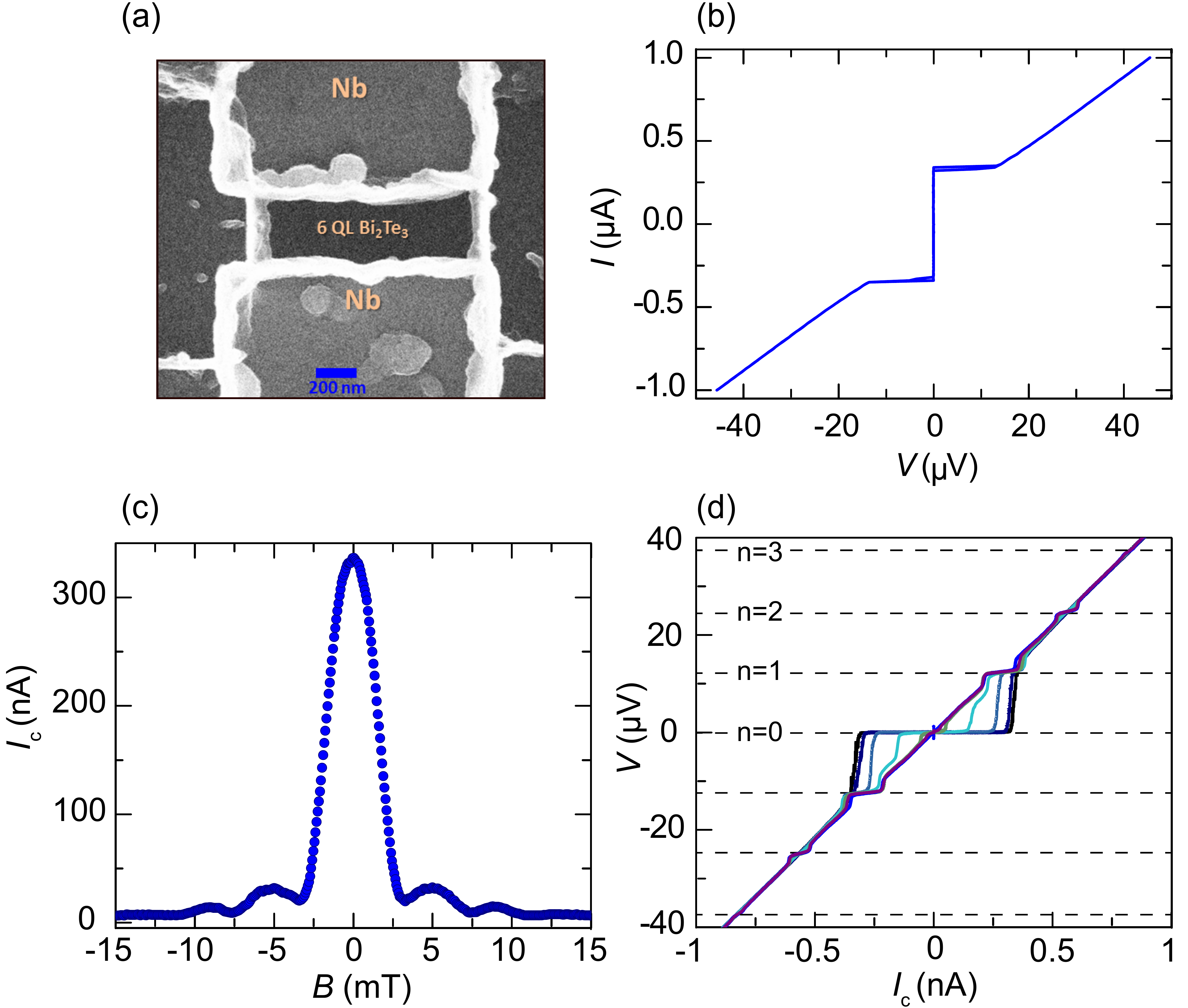}
	\caption{Josephson junction characteristics of Nb/\BiTe/Nb devices. (a) A scanning-electron micrograph image of a typical Josephson junction device. The device dimensions (width: 860 nm and length: 250 nm) are clearly visible by residues of cross-linked photoresist from the etch mask. (b) The I-V characteristic of a Josephson device with 6~nm thick \BiTe~weak link (JJ6/1), and (c) a representative Fraunhofer-like Josephson diffraction pattern. (d) Voltage plateaus appear in the V-I curves under microwave irradiation with a frequency of 6.02~GHz. The voltages $V_n \simeq n \times (12.4~\mu$V) for Shapiro steps of order $n$ are indicated for several RF drive amplitudes. The Josephson junction characteristics in (b)-(d) were measured at 15 mK.}
	\label{FIG:JJ}
\end{figure*}
\section*{Sample preparation and Josephson junctions characteristics}
\paragraph*{}High-quality thin films of \BiTe\, were
grown by molecular beam epitaxy (MBE). Hall bar devices and Josephson junctions were patterned side-by-side using standard electron-beam lithography, and subsequent dry etching and sputter deposition of electrodes. Further details on the growth procedure of the thin films and device fabrications are presented in Supplemental Material and ref.~\cite{ngabonziza_situ_2018,ngabonziza_situ_2015}. Figure~\ref{FIG:JJ}\textcolor{blue}{(a)} shows a typical Nb/\BiTe/Nb Josephson junction (JJ6/1).  The weak link is 860~nm wide and 250~nm long.  The device was patterned on a 6~nm-thick film of \BiTe\, which was grown on a (111) oriented {Sr}{Ti}{O}$_3$ (STO) substrate. The current-voltage characteristic (IVC) is plotted in Figure~\ref{FIG:JJ}\textcolor{blue}{(b)}.  The device exhibits sharp switching into the voltage state with little hysteresis.  It displays a Fraunhofer-like Josephson diffraction pattern with perpendicular applied magnetic field [Figure~\ref{FIG:JJ}\textcolor{blue}{(c)}], which indicates a uniform critical current density in the weak link.  When irradiated with microwaves of frequency $f = 6.02$~GHz, the IVC develops voltage plateaus\textsuperscript{\cite{shapiro_josephson_1963}} corresponding to multiples $n$ of the driving frequency $f$, $V_n = \frac{n h f}{2 e} \simeq n \times (12.4~\mu$V), where $h$ and $e$ are Planck's constant and the elementary charge, respectively [Figure~\ref{FIG:JJ}\textcolor{blue}{(d)}]. Both, even-$n$ and odd-$n$ steps, are present in these Josephson devices.
\section*{Gate dependence of Josephson effect and charge carrier density}
\begin{figure*}[t!]
	\includegraphics[width=0.925\textwidth]{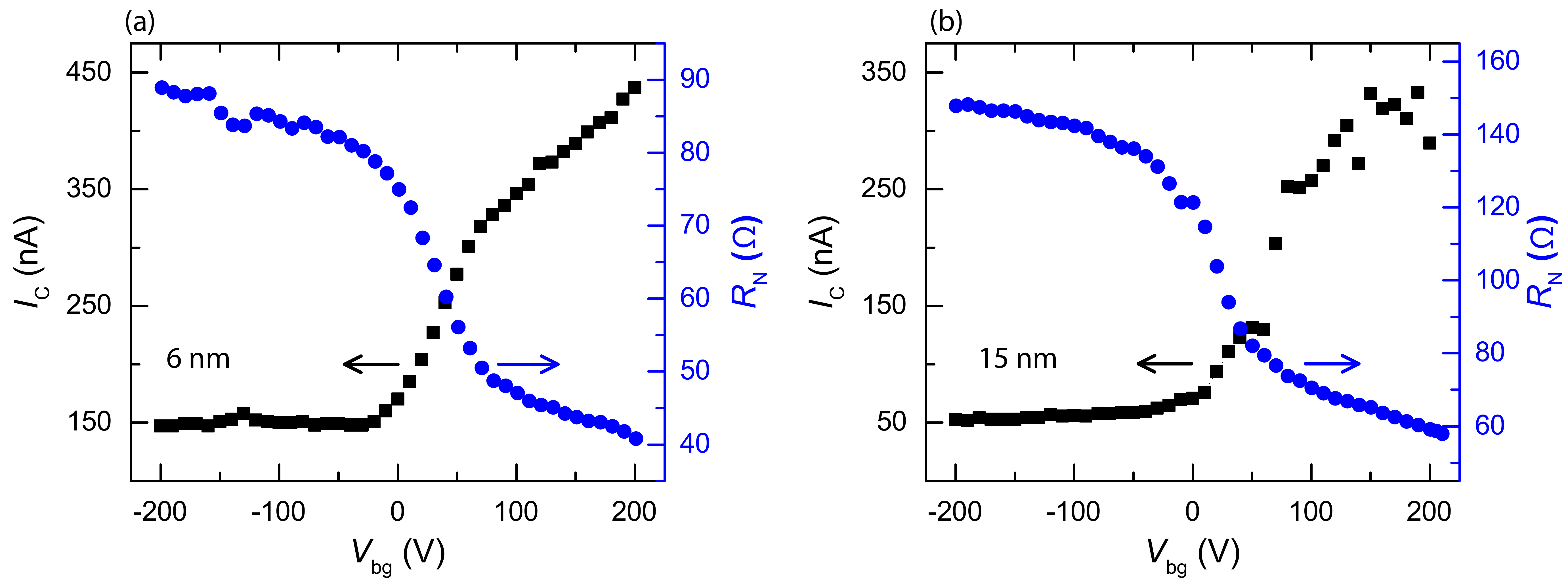}
	\caption{Back-gate voltage dependence of the normal state resistance and the Josephson critical current measured at 15 mK for Nb/\BiTe/Nb junctions of (a) 6~nm (JJ6/1) and (b) 15~nm (JJ15/1) thickness of the  \BiTe\ weak link. We find that the critical current is enhanced for positive back-gate voltages, it flattens out when negative gate bias is applied. For both devices, the normal state resistance R$_\text{N}$ drops by a factor of $\gtrsim 2.2$ with increasing gate voltage.}
	\label{FIG:IcVbg}
\end{figure*}

\paragraph*{}
We have modulated the charge carrier distributions in the films by electrostatic gating. We have used the STO substrate as the back-gate dielectric since it has a high dielectric constant $( \epsilon_\mathrm{STO} \approx 2-6 \times 10^4)$ at low temperature.\textsuperscript{\cite{goryachev_determination_2015}}   Voltages $V_\mathrm{bg}$ in the range of $\pm 200$\,V are sufficient to create a triangular quantum well or to deplete carriers in the bottom region of the film.\textsuperscript{\cite{ngabonziza_situ_2016}}  Figure~\ref{FIG:IcVbg}\textcolor{blue}{(a)} and Figure~\ref{FIG:IcVbg}\textcolor{blue}{(b)} depict the effect of charge (re-)distribution in the material on the Josephson effect for two representative Josephson junctions that were patterned on films of 6~nm (JJ6/1) and 15~nm (JJ15/1), respectively.  At  $200$~V gate bias, the Josephson coupling energy amounts to $e I_\mathrm{c} R_N \approx 18~\mu$eV for both devices.  The Josephson critical current $I_c$ is defined by the voltage criterion $V \leq 1~\mu$V. An estimate of the normal state resistance of the device is obtained from a fit to the IVC at large current bias ($> 1.5~\mu$A).  The gating characteristics of the devices are identical.  At large negative gate bias, $I_c$ is approximately constant.  For positive gate voltage, it rises sharply at first, then grows with a smaller slope at high bias.  In this region, device JJ15/1 shows larger hysteresis and stochastic switching [Figure~\ref{FIG:IcVbg}\textcolor{blue}{(b)}].  For the values of $I_c$ and $R_N$, the hysteresis is expected to arise from the phase dynamics,\textsuperscript{\cite{antonenko_quantum_2015}} not the (geometric) capacitance of the weak link.  The normal state resistance $R_N$ drops monotonously with increasing gate voltage.  In the transition region around $V_{\rm{bg}} = 0$, the decrease is more prominent.  
\paragraph*{}  Next, we compare the gating behavior of $I_c$ with the evolution of the charge carrier densities in the film.  We have performed magnetoresistance measurements as a function of back-gate voltage on a Hall bar device fabricated on the same 15~nm-thick film as the JJ15/1 device for a sample temperature of 50 mK. The Hall data were fitted with a standard two-carrier model expression for $R_{xy}$ using the zero-field sheet resistance $R_S$ as a constraint [Figure~\ref{fig:S1}\textcolor{blue}{(a)} in Supplementary Material]. Figure~\ref{FIG:MR}\textcolor{blue}{(a)} [black filled squares] gives the total carrier density $n$ as function of gate voltage. Similar to the critical current data, saturation at $2.2 \times 10^{13}\,\mathrm{cm}^{-2}$ in the depletion region  was observed and a carrier density increase for positive gate bias with a slope change around 60~V and a maximum value of $5.3 \times 10^{13}\,\mathrm{cm}^{-2}$ at 200~V. The mobilities of the two carrier types change only slightly with gate voltage.  The extracted high- and low-mobility carriers are $\sim 1900~\mathrm{cm}^2 /\mathrm{V.s}$ and $\sim 700~\mathrm{cm}^2 /\mathrm{V.s}$, respectively.
\begin{figure*}[!t]
	\includegraphics[width=0.925\textwidth]{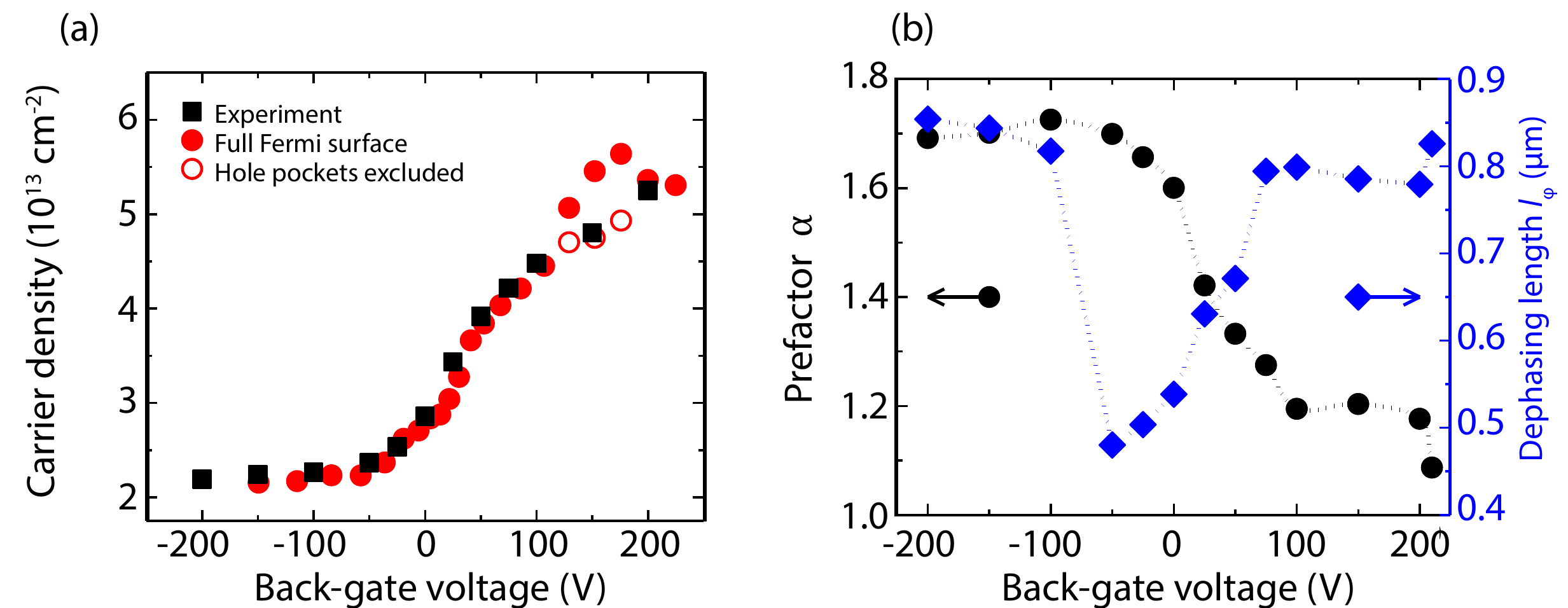}
	\caption{Carrier density, weak antilocalization fit data, and modeling of the carrier distribution in a 15~nm-thick \BiTe\, film.  (a) The total carrier density as a function of gate voltage from Hall measurements [black filled square], and from band structure modeling [red filled circles]. The open red circles at high voltages are obtained by excluding disconnected hole pockets close to the $M$ point of the band structure from the summation (b) Prefactor $\alpha$ and dephasing length $l_\phi$  from a fit to the weak antilocalization correction of the conductance.  A clear increase in the magnitude of the weak localization signal indicates the decoupling of the bottom TSS from the bulk.}
	\label{FIG:MR}
\end{figure*}
\section*{Self-consistent band structure calculations}
\paragraph*{} To understand the charge distribution in the films at different gate voltages, we have modeled the band structure of a 15~nm-thick film of \BiTe\, using a tight-binding Hamiltonian with parameters from ref.\cite{lee_tight-binding_2006}. To determine the correct doping level, a series of solutions to the coupled Schr\"odinger and Poisson equations were calculated self-consistently using the NEMO5 software package.\textsuperscript{\cite{steiger_nemo5:_2011}}  For each series, the chemical potential and the top surface electrostatic potential were kept fixed and the bottom electrostatic potential was varied.  Then, we have calculated the carrier density at the Fermi level and the (approximate) back-gate voltage. Representative cuts of the band structure between points of high symmetry along the line $\mathrm{K} - \Gamma - \mathrm{M}$ are plotted in Figure~\ref{fig:S2}\textcolor{blue}{(a)}-\textcolor{blue}{(c)} (for details on calculations, see Supplementary Materials). Figure~\ref{FIG:MR}\textcolor{blue}{(a)} shows good agreement between experimental data [black filled squares] and extracted carrier densities of the model calculation [red filled circles]
for a chemical potential shift of 300 meV and a top surface electric voltage of -0.3~V.  These parameters concur with results of ARPES measurements on \BiTe\, films of the same growth series with identical growth parameters.\textsuperscript{\cite{ngabonziza_situ_2015}}  The discrepancy at higher gate voltages is resolved by excluding disconnected hole pockets near the M-point of the Brillouin zone from the carrier density estimate [open circles in Figure~\ref{FIG:MR}\textcolor{blue}{(a)}]. We have found that the presence of TSS makes it difficult to deplete transport carriers, it is easy to enlarge the Fermi surface by applying a small positive gate bias.

\paragraph*{}  Electrostatic gating changes the shape of the confining potential and the distribution of transport carriers in the well.  Consequently, as a function of gate bias, different sections of the film participate in transport.  Depending on the effective scattering length of the defect potentials, the disorder potential landscape of each section may be different.  The scattering rate between the sections depends on the wavefunction overlap. We have observed the effect experimentally as a gradual increase in the weak antilocalization (WAL) correction to the longitudinal resistance of the Hall bar structures.  To quantify the change, we have fitted the magnetoconductance data [Figure~\ref{fig:S1}\textcolor{blue}{(b)} in Supplementary Materials] to the  Hikami-Larkin-Nagaoka~(HLN) formula:\textsuperscript{\cite{hikami_spin-orbit_1980, maekawa_effects_1981}} 
\begin{equation}
G - G_b = \alpha \frac{e^2}{2 \pi \hbar} \Big[ \ln \Big( \frac{\hbar}{4 e l_\varphi^2 B} \Big) - \Psi \Big(\frac{1}{2} + \frac{\hbar}{4 e l_\varphi^2 B} \Big) \Big]\ ,
\end{equation} 
after subtracting a quadratic background $G_b$. Here, $\Psi$ denotes the digamma function. Figure~\ref{FIG:MR}\textcolor{blue}{(b)} depicts the back-gate voltage dependence of prefactor~$\alpha$ and the dephasing length $l_\varphi$, quantities that characterize the magnitude of the correction and the length scale of electronic phase coherence, respectively. The dephasing length was found to be $\sim 0.8~\mu$m to either side of the transition but dips to $\sim 0.48~\mu$m in between. At large positive gate bias, the prefactor $\alpha$ is $ \sim 1.1$ and increases to $\sim 1.7$ at large negative gate voltage. The transition occurs gradually in the same bias region where we have observed a decrease in the Josephson critical current. 

\paragraph*{} The spatial redistribution of charge carriers allows for a partial decoupling of transport on the bottom surface.  Naively, one might expect an increase in the WAL signal as scattering between the TSS and the bulk is suppressed, and a separate conduction channel forms.  A semiquantitative interpretation of the magnitudes of the correction, however, requires a detailed analysis of the number of Cooperon modes (interference contributions to weak (anti-)localization) in the thin film.\textsuperscript{\cite{garate_weak_2012}} The two relevant limits are: Firstly, a system of coupled TSS and bulk quantum well states (QWS) in the limit of dephasing time much larger than spin and valley (`top' or `bottom') scattering time\textsuperscript{\cite{footnote}} for which two (hybridized) spin-singlet bulk Cooperon modes exist ($e^2/h$ correction). This system corresponds to the sample at large positive gate voltages. Secondly, when the bottom TSS is only weakly coupled to the QWS, a third Cooperon mode from the topological surface contributes ($\frac{3}{2}\,e^2/h$ correction), explaining the observed increase of the weak antilocalization signal for sample depletion. 
The smooth transition between the two regimes is governed by the characteristic resistances of the bulk QWS and TSS channels. They are set by the dephasing lengths in the respective channels which change with the hybridization of Cooperon modes. Decoupling of the TSS is then observed as a rebound in the dephasing length to its larger, initial value at the lowest gate voltages [Figure~\ref{FIG:MR}\textcolor{blue}{(b)}].  The experimental data are in good qualitative agreement with the theoretical scenario.\textsuperscript{\cite{garate_weak_2012}}  Deviations may arise, e.g., from additional corrections which are specific to topological surface states and modify the HLN expression.\textsuperscript{\cite{GTkachov_2015}}
\begin{figure*}[t!]
	\includegraphics*[width=0.925\textwidth]{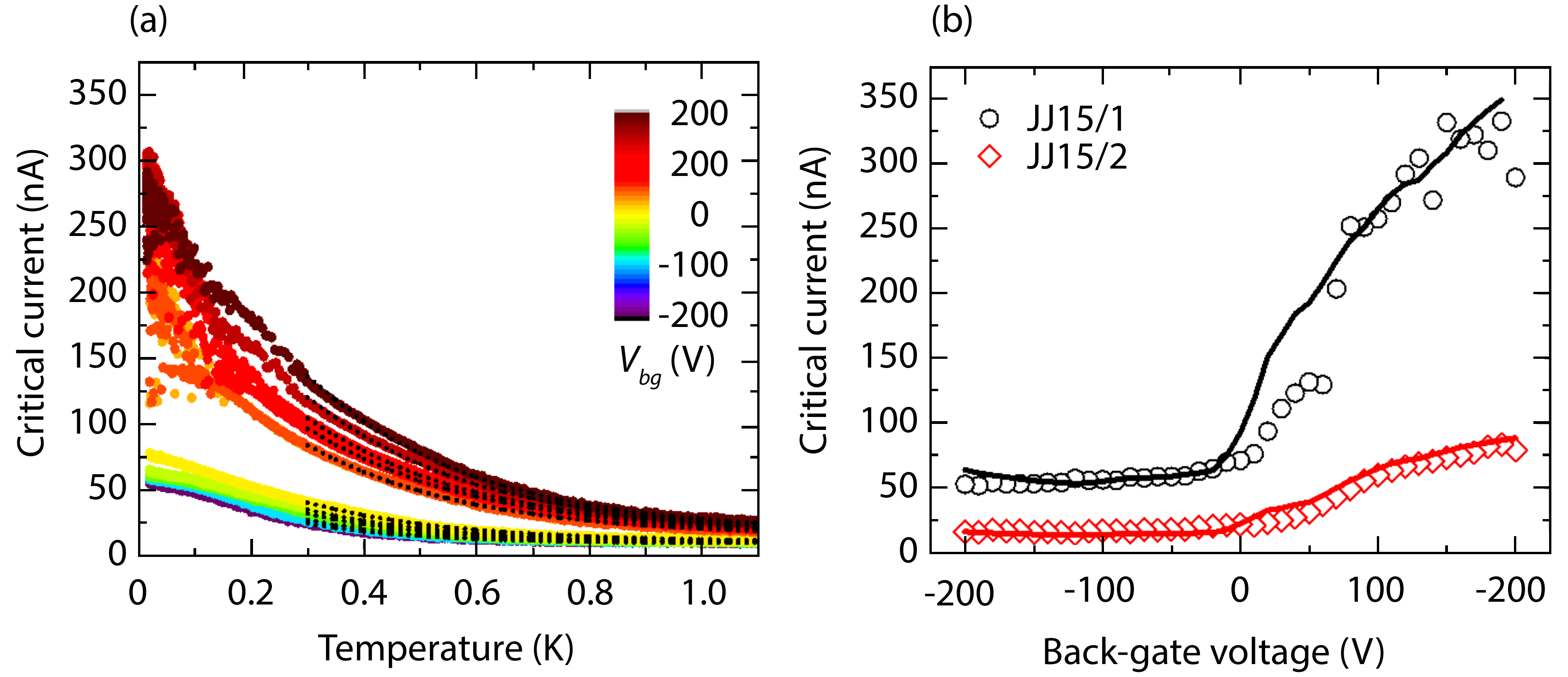}
	\caption{Critical current of Josephson devices with 15 nm-thick \BiTe weak links. (a) Temperature dependence of the critical current in JJ15/1 for different gate voltages.  A fit to the high temperature data was performed to estimate the magnitude of the Thouless energy [dashed lines]. (b) Critical current of devices JJ15/1 and JJ15/2 at different gate voltages.  Open symbols represent experimental values.  Solid lines are obtained from band structure data using a single scaling factor for the entire gate voltage range.  Stochastic switching sets in above $\sim 120$~nA.}
	\label{FIG:Icfit}
\end{figure*}
\paragraph*{} Finally, using the band structure model of the thin film, we could understand the strong enhancement of the Josephson critical current that surpasses the increase in charge carriers multiple times. 
As all Josephson devices have $eI_cR_N$-products significantly smaller than the bulk pairing potential in the Nb electrodes ($\Delta \approx 1.3$~meV; estimated from the critical temperature of the Nb electrodes), and $I_c$ varies slowly with temperature [Figure~\ref{FIG:Icfit}\textcolor{blue}{(a)}], the devices are SNS contacts in the diffusive, long junction regime.  The electronic phase coherence limits the critical current, and the scale of the Josephson energy is set by the Thouless energy ($E_C$) of electrons traversing the weak link: $e I_c R_N = \gamma E_C$.
In ordinary metallic wires,\textsuperscript{\cite{dubos_josephson_2001, dubos_coherent_2001}} the prefactor $\gamma$ was found to be close to the theoretical value,\textsuperscript{\cite{wilhelm_supercurrent_1997}} $\gamma = 10.82$. Using graphene as gate-tunable weak link material, the scaling relationship was demonstrated to hold over more than two orders of magnitude in $E_C$,\textsuperscript{\cite{li_full_2016, ke_critical_2016}} although with a strongly reduced prefactor.    
We have established the magnitude of the Thouless energy in the system by fitting the $I_c-T$ curves in Figure~\ref{FIG:Icfit}\textcolor{blue}{(a)} with the expression for long diffusive contacts:\textsuperscript{\cite{dubos_josephson_2001}}
\begin{equation}\label{eq:IcT}
e I_c R_N \propto E_C^\star \left[ \frac{L}{L_T}\right] \exp \left(- \frac{L}{L_T}\right)\,,
\end{equation}
valid for $T\gg 5 E_C^\star/k_\mr{B}$. Here, $L_T$ is the thermal 
coherence length $(L_T = \sqrt{\hbar D/2\pi k_\mr{B}T})$ and $L$ the length of the junction. We have obtained values of $E_C^\star = 4.3 - 6.6\,\mu$eV. To understand the gating behavior of $I_c$, we have calculated the average diffusion constant $\overline{D}$ from the model band structures (for details on calculations, see Supplementary Materials).  
Figure~\ref{FIG:Icfit}\textcolor{blue}{(b)} [solid lines] shows calculated critical current for two junctions with 15 nm-thick \BiTe~ weak links.  The shapes of the $I_c-V_\mr{bg}$ curves are very well reproduced.  The scaling factor is two orders of magnitude larger than the theoretical value in ref.~\cite{wilhelm_supercurrent_1997},  but it compares well to the reported suppression of critical current in graphene Josephson devices.\textsuperscript{\cite{li_full_2016}}  However, we provide a different interpretation.  In ref.~\cite{li_full_2016}, the discrepancy is attributed to the interface resistance between the electrodes and the graphene.  For devices presented in this work, the contact resistance is estimated to be about 20~\% of $R_N$.  The reduction in $I_c$ should thus be attributed to additional scattering centers due to the damage incurred during device fabrication.  Unlike graphene, the added defects facilitate scattering between ca. 20 bands hence the significant increase in scattering frequency and a smaller critical current.  In our experiments, there is no indication of the decoupled bottom surface contributing to the supercurrent transport.  A possible explanation is that the carrier density on the bottom surface is very small compared to the bulk subbands.  Further, the depletion zone acts as additional barrier for Cooper pair tunneling.
\section*{Conclusion}
 We have studied the gate-dependence of the Josephson effect, the carrier densities, and the weak antilocalization correction to the conductivity in devices fabricated from thin \BiTe\, films.  While the carrier densities were enhanced by applying positive gate voltages, the samples could not be depleted.  This was explained by calculating the carrier distribution in the thin film self-consistently using a tight-binding model.  The theoretical calculation has reproduced the experimental values for the transport carrier densities well.  While positive gate voltages created a potential well, negative gate bias formed a depletion zone in the material.  This scenario was corroborated by the analysis of the WAL signal in magnetotransport measurements. The Josephson devices were found to be in limit of long, diffusive SNS junctions, and the critical current scaled with the estimated Thouless energy obtained from a calculation of the average diffusion constant in the band structure model. 

 We conclude that a satisfactory description of all transport experiment on the \BiTe\,thin film devices could be found by taking the carrier distribution in the sample and the shape of the subbands into account.  The evolution of the carrier density with applied gate voltage followed naturally from the profile of the electrostatic potential in the film, and no \textit{ad hoc} assumptions about the charge distribution in the sample or the capacitance of surface and bulk states were required.  In particular, we want to stress that for a careful evaluation of the WAL signal, we cannot simply interpret it as a sum of  topological surface states and bulk, but must look at the probabilities of scattering between different two-dimensional electron systems and the resulting corrections to the conductivity.  Ultimately, from a careful, systematic analysis, we are able to reveal that the critical current in TI Josephson devices maps band structure properties of the thin films. These data thus provide a coherent picture that incorporates band bending in the discussion of proximity effects for 3D TI devices. We like to finish by stressing that this analysis could be extended to discuss proximity superconductivity in other TI and narrow gap semiconductor systems for which band bending plays a role.
  \\ \\
This work was financially supported by the Netherlands
Organization for Scientific Research (NWO) and the European Research Council (ERC) through a Consolidator Grant.
 \newpage

\begin{thebibliography}{99}
\bibitem{footnote}In the present context, the spin scattering time characterizes dephasing by scattering with ordinary impurities when the spin-orbit interaction is strong, not spin-flip scattering from magnetic impurities. Valley scattering refers to scattering between the `top' and `bottom' sectors of the Hamiltonian for (massive) Dirac fermions.
\end{thebibliography}

\onecolumngrid
\newpage
\setcounter{table}{0}
\setcounter{figure}{0}
\renewcommand{\thefigure}{S\arabic{figure}}%
\setcounter{equation}{0}
\renewcommand{\theequation}{S\arabic{equation}}%
\setstretch{1.5}
\begin{center}
\title*{\textbf{\Large{Supplementary Information:} \\ [0.25in] \large{{Josephson Effect and Charge Distribution in Thin \BiTe\ Topological Insulators}}}}
\end{center}
\begin{center}
\Large{M.P. Stehno$^{\textcolor{blue}{\small{*}}}$}\\
\small{Faculty of Science and Technology and MESA$^+$ Institute for Nanotechnology, University of Twente, 7500 AE Enschede, The Netherlands \\
Physikalisches Institut EP3, University of W\"urzburg, Am Hubland, D-97070 W\"urzburg}
\end{center}

\begin{center}
\Large{P. Ngabonziza$^{\textcolor{blue}{\small{*}}}$}\\
\small{Faculty of Science and Technology and MESA$^+$ Institute for Nanotechnology, University of Twente, 7500 AE Enschede, The Netherlands\\
Max Planck Institute for Solid State Research, D-70569 Stuttgart, Germany\\
Department of Physics, University of Johannesburg, P.O. Box 524 Auckland Park 2006, Johannesburg,\\ South Africa}\\
\end{center}

\begin{center}
\Large{H. Myoren}\\
\small{Graduate School of Science and Engineering, Saitama University, 255 Shimo-Okubo, Sakura-ku, Saitama 338-8570, Japan}
\end{center}

\begin{center}
\Large{A. Brinkman}\\
\small{Faculty of Science and Technology and MESA$^+$ Institute for Nanotechnology, University of Twente, 7500 AE Enschede, The Netherlands}
\end{center}
\section*{\large{S\lowercase{ample} P\lowercase{reparation and} D\lowercase{evice} F\lowercase{abrication}}}
\paragraph*{} Recent improvements in the molecular-beam epitaxy (MBE) of topological insulator materials have allowed us to obtain high-quality thin films of \BiTe\, with low intrinsic doping.\textsuperscript{\cite{ngabonziza_situ_2015, hoefer_intrinsic_2014}} 
We have grown \BiTe\, films of 6~nm and 15~nm (average) thickness on (111) oriented {Sr}{Ti}{O}$_3$ substrates  using a two-step deposition process.\textsuperscript{\cite{ngabonziza_situ_2015,ngabonziza_situ_2018}}  The films are thick enough to avoid opening a large hybridization gap in the band structure.\textsuperscript{\cite{zhang_crossover_2010, liu_oscillatory_2010}}
 They feature a granular morphology with single-crystalline grains of $\gtrsim 300\,\mathrm{nm}$ lateral dimension. By varying the substrate temperature, we have control over the intrinsic doping of the film\textsuperscript{\cite{wang_topological_2011}} and place the Fermi level close to the bottom of the conduction band, about $300$\, meV above the Dirac point. This is verified routinely by \textit{in-situ} angle-resolved photoemission spectroscopy (ARPES), see 
 ref.~\cite{ngabonziza_situ_2015,ngabonziza_situ_2018}.  Unlike {Bi}$_2${Se}$_3$, the material is not prone to forming surface vacancies (cp. ref.~\cite{PDCKing2011,MatthewBrahlek2015,bansal_thickness-independent_2012,bianchi_coexistence_2010}).  The Fermi level in ARPES measurements thus corresponds roughly to the electrochemical potential of the film as the film thickness is comparable to the electrostatic screening length ($\gtrsim 10$\,nm at typical carrier densities of $\sim  10^{13}$~{cm}$^{-2}$).\textsuperscript{\cite{MatthewBrahlek2015, jenkins_dirac_2013-1}}

Hall bar devices and Josephson junctions were patterned side-by-side using standard electron-beam lithography and sputter deposition of 60~nm thick Nb electrodes with a 5~nm Pd capping layer after cleaning the contact area with a low-power Ar plasma.  The Hall bars and the weak link areas were shaped by 
dry etching.  To minimize the exposure to chemicals, we have not removed the cross-linked resist residues [outline in Figure~\ref{FIG:JJ}\textcolor{blue}{(a)} in the Main Text].  The device boundaries were shaped by dry etching. The samples were mounted on a printed-circuit board using silver paint to ensure electrical contact between the crystalline substrate and the backgate electrode. Electrical transport characterizations were carried out in a dilution refrigerator with heavily filtered signal lines.
\section*{\large{M\lowercase{agnetoconductance} M\lowercase{easurements and} T\lowercase{wo-band} F\lowercase{its}}}
\label{SUP:MR}
We have measured the magnetotransport properties of \BiTe\ films at different back-gate voltages for a sample temperature of 50 mK. The measurements were performed in Hall bar geometry using standard lock-in techniques. The lateral dimensions of the Hall bar were: $L=250\ \mu \text{m}$, and $W= 50\ \mu \text{m}$. A small excitation current of $5$ nA was chosen to minimize sample heating. Here, we present magnetotransport data of the representative sample of 15 nm-thick \BiTe\ film (same sample discussed in the Main Text).
\begin{figure}[!b]
	\includegraphics[width=1\textwidth]{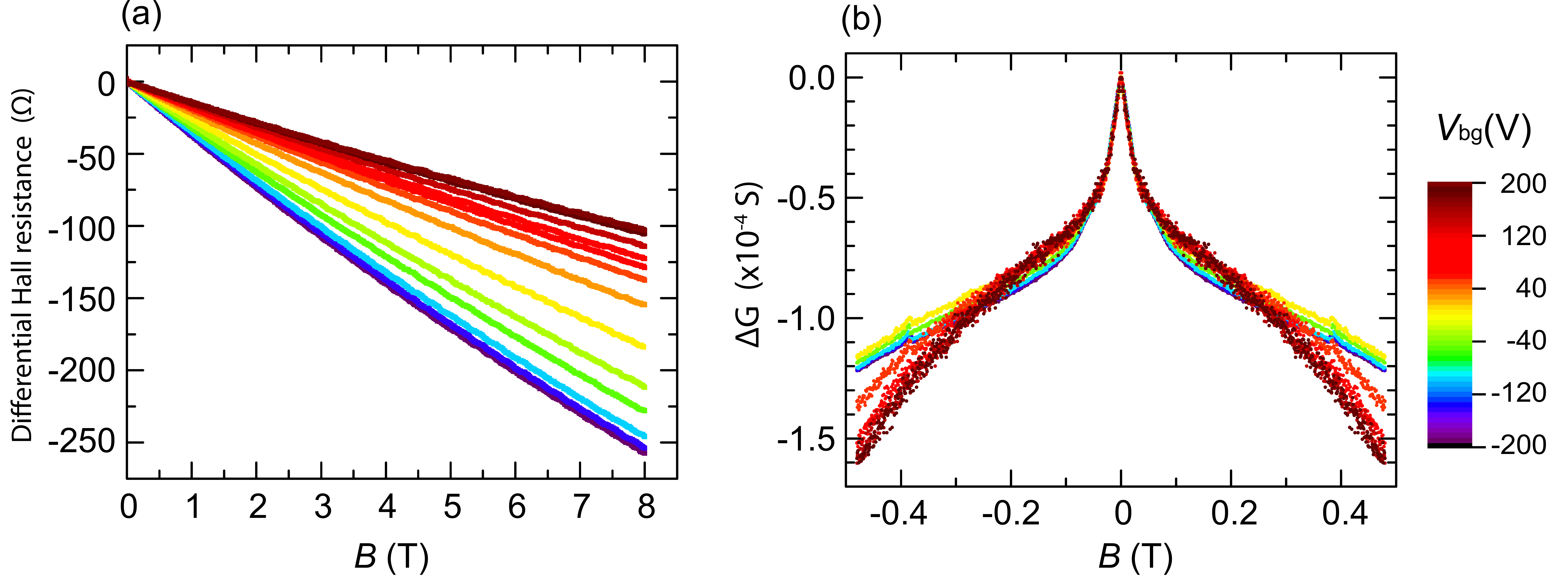}
	\caption{Magentoconductance data of the 15~nm-thick film (a) Antisymmetrized Hall resistance as a function of back-gate voltage. A small nonlinearity indicates multiband conduction. (b) The magnitude of the WAL correction and the shape of $\Delta$G change with the applied back-gate bias. }
	\label{fig:S1}
\end{figure}
We have fitted the (antisymmetrized) Hall resistivity ($R_{xy}$) [Figure~\ref{fig:S1}\textcolor{blue}{(a)}] with a standard two-band model expression,
\begin{equation}
R_\mathrm{xy}=-\Big(\frac{B}{e}\Big)\frac{n_1\mu_1^2 + n_2\mu_2^2 + B^2 \mu_1^2 \mu_2^2 (n_1+n_2)}{(|n_1|\mu_1 + |n_2|\mu_2)^2+B^2 \mu_1^2 \mu_2^2 (n_1+n_2)^2}\ ,
\end{equation} 
with the constraint that the conductance must equal the inverse of the sheet resistance, $R_S^{-1} = e\sum_i n_i \mu_i$. Here, $n_i$ and $\mu_i$ denote sheet carrier density and mobility of carrier type $i$, respectively. 
Additionally, we have chosen to minimize the carrier density of the high-mobility carriers (HMC) to further restrict the fit parameter range. In this way, the carrier densities of the two bands follow the estimates for surface and bulk carrier densities in the self-consistent model closely. However, the value of the band-averaged diffusion constant changes only slightly when the second condition is relaxed. 

The sheet resistance showed a weak antilocalization (WAL) feature at low magnetic fields ($B\lesssim 200$ mT). After (anti-)symmetrization and inversion of the resistivity matrix, we have subtracted the conductance at zero field to obtain the conductance $\Delta G (B)$ at different back-gate voltages [Figure~\ref{fig:S1}\textcolor{blue}{(b)}]. To quantify the WAL correction, we have fitted the sheet conductance with the Hikami-Larkin-Nagaoka (HLN) expression. Detailed discussions on the extracted prefactor $\alpha$ and dephasing length $l_{\varphi}$ at different back-gate biases are presented in the Main Text.

\section*{\large{S\lowercase{elf-consistent} B\lowercase{and} S\lowercase{tructure} C\lowercase{alculations}}}
\begin{figure}[t!]
	\includegraphics[width=1\textwidth]{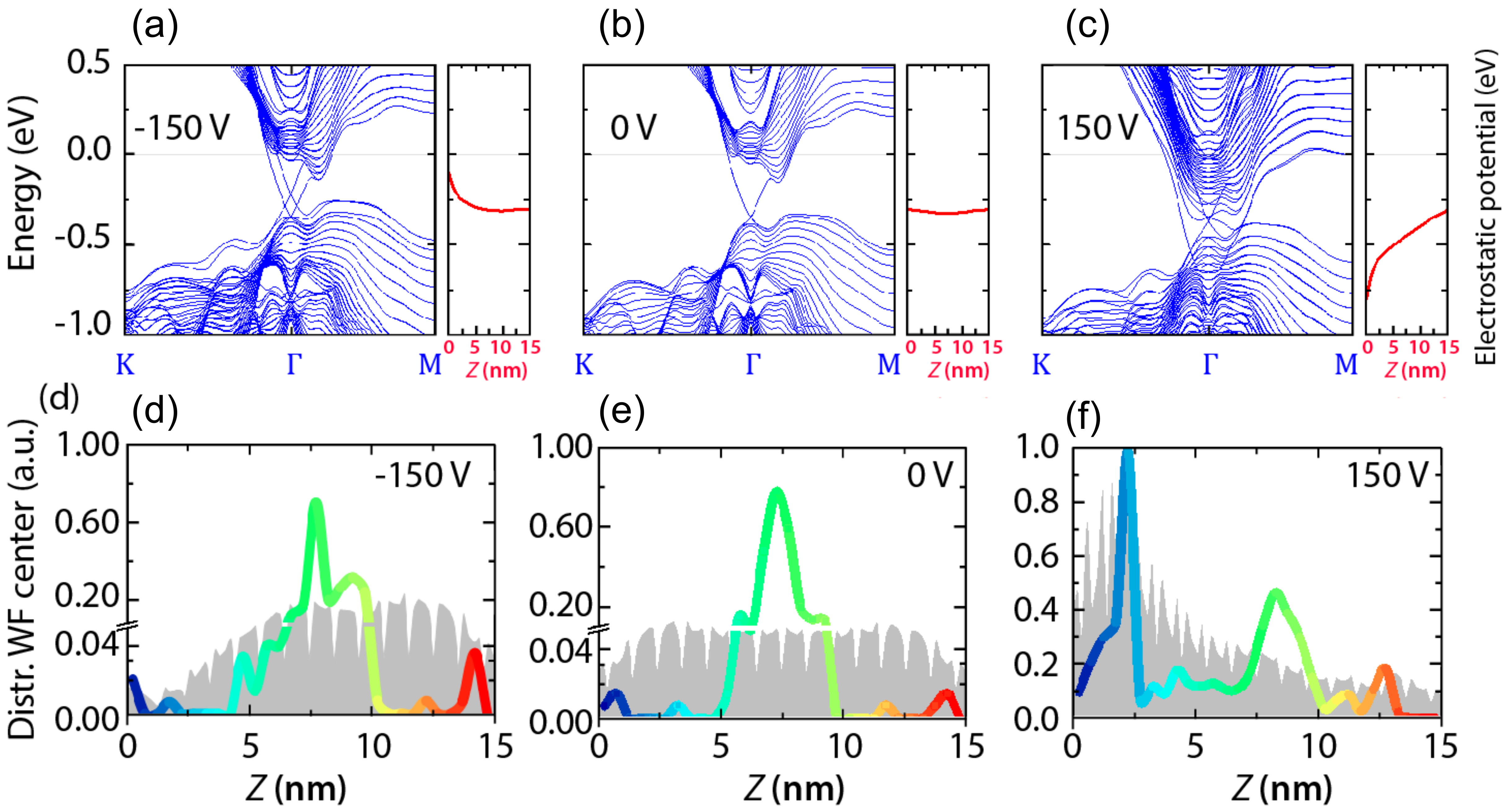}
	\caption{Calculated band structure plots and electrostatic potential profiles (red curves) in the 15~nm-thick film at (a) -150 V, (b) 0 V and (c) 150 V back-gate voltages.  Up to 26 bulk (sub)bands are populated.  Charge distribution  of carriers at (d) -150 V, (e) 0 V and (f) 150 V back-gate voltages for the Fermi level located at distance $z$ from the bottom interface of the topological insulator material [gray-shaded area] and the median of the probability density [bold trace].}
	\label{fig:S2}
\end{figure}

Band structure calculations were carried out for 15~nm-thick film of \BiTe\, using a tight-binding Hamiltonian with parameters from ref.~\cite{lee_tight-binding_2006}. The correct doping level and boundary conditions were emulated by introducing a chemical potential shift and fixing the surface electric potentials. The solutions of the coupled Schr\"odinger and Poisson equations were calculated self-consistently using the NEMO5 software package.\textsuperscript{\cite{steiger_nemo5:_2011}} 
Figures~\ref{fig:S2}\textcolor{blue}{(a)}-\textcolor{blue}{(c)} gives calculated band structure plots and electrostatic potential profiles at different back-gate biases. A constant chemical potential shift of 300 m$\rm{eV}$ 
and a surface voltage of -0.3~V led to best agreement with experiment carrier density data [Figure~\ref{FIG:MR}\textcolor{blue}{(a)} in the Main Text].  These parameters agree with ARPES experiments on \BiTe\, films prepared in the same growth series under similar conditions.\textsuperscript{\cite{ngabonziza_situ_2015}}  Here, we have also assumed few charged surface contaminants on the film surface.  The bottom surface voltage was varied between 0 and -1\,V. We have chosen  $\epsilon_\mathrm{TI}=75$ for the dielectric constant of \BiTe.\textsuperscript{\cite{richter_raman_1977}} The band structure data was then sampled in a narrow energy interval ($\Delta E = 20\,\mathrm{meV}$) around the position of the electrochemical potential in the film ($E = 0$).  The number of charge carriers was found by calculating a weighted sum over this section of the band structure (taking into account the area element associated with each reciprocal space vector) and rescaling the result to the width of the Fermi distribution at the measurement temperature $T = 100\,\mathrm{mK}$. The back-gate voltage was calculated using electrostatic boundary conditions and the electric field profile in \BiTe\, from the  solution of the Poisson equation. Since it is difficult to measure the dielectric constant of STO~[111] precisely, we kept $\epsilon_\mathrm{STO}$ as a variable and determined its value by scaling the $V_\mathrm{bg}$ axis to match the experimental data. This yields $\epsilon_\mathrm{STO} = 35,000$, a value slightly lower than expected (cp. ref.~\cite{goryachev_determination_2015}). The numeric result for the sheet carrier density is plotted in Figure~\ref{FIG:MR}\textcolor{blue}{(a)} [red filled circles]. A discrepancy appears to exist around 150~V which could be resolved by excluding disconnected hole pockets close to the $M$ point of the band structure from the summation [open red circles in Figure~\ref{FIG:MR}\textcolor{blue}{(a)}]. Such hole pockets originate from bands that warp down close to the Brillouin zone edge [Figure~\ref{fig:S2}\textcolor{blue}{(a)}-\textcolor{blue}{(c)} right panel].   With this correction, we have obtained excellent agreement between the two-carrier model fits and the band-structure-based calculation of the sheet carrier density. Figures~\ref{fig:S2}\textcolor{blue}{(d)}-\textcolor{blue}{(f)} gives the charge distribution  of carriers at different back-gate voltages for the Fermi level located at distance $z$ from the bottom interface of the topological insulator material [gray-shaded area] and the median of the probability density [bold trace]. Whereas the bottom topological surface state (blue) decouples from the bulk at large negative gate voltages, the carriers could scatter freely between the top surface and bulk otherwise. 

To analyze the gating behavior of $I_c$, we have extracted the average diffusion constant $\overline{D}$ from the model band structures.  Here, $\overline{D} = \frac{1}{2} \sum_{i}{n_i \overline{v}_{F,i}^2 \tau_i}/\sum_{i}{n_i}$ is a weighted average over surface band and bulk (sub)bands where $\overline{v}_{F,i}$ are the averaged Fermi velocities.  Information about the scattering dynamics are encoded in the transport relaxation times, which are calculated from the mobilities $\mu_i$ and the (mean) effective band masses $m^*_i$ by the standard relation $\tau_i = m^*_i \mu_i/e$.  As the gate dependence of the bulk and surface carrier densities in the band structure calculation closely follows the experimental values obtained for the two carrier types in the two-band model fits of the magnetotransport data, we have associated the two mobility values with bulk and surface carriers, respectively.  Values for missing gate voltages were obtained by linear interpolation.  The Thouless energy is given by $E_C=\hbar \overline{D}/L^2$.  The result is rescaled by a constant factor $\gamma^\star$  to obtain an estimate for the critical current using the relation $e I_c R_N =\gamma^\star E_C$. Using this procedure, the shape of the measured $I_c - V_{bg}$ was very well reproduced.

\end{document}